\documentclass[ams, prb, twocolumn, amssymb,floatfix]{revtex4-1}
\usepackage{amsmath}
\usepackage{tabularx}
\usepackage{bm}
\usepackage{euscript}
\usepackage{graphicx}
\usepackage{color}
\usepackage{amsfonts}
\usepackage{exscale}
\usepackage{amsbsy}
\usepackage{subfigure}
\usepackage{textcomp}
\usepackage{comment}
\usepackage{hyperref}

\newcommand{\ket}[1]{\left|#1\right\rangle}
\newcommand{\bra}[1]{\left\langle#1\right|}

\newcommand{\nn}{\nonumber}

\newcommand{\beq}{\begin{equation}}
\newcommand{\eeq}{\end{equation}}
\newcommand{\be}{\begin{eqnarray}}
\newcommand{\ee}{\end{eqnarray}}

\begin{document}

\title{ Why is the HLR theory particle-hole symmetric?}
\author{Prashant Kumar$^{1}$, Michael Mulligan$^{2}$, S. Raghu$^{1,3}$}
\affiliation{$^{1}$Stanford Institute for Theoretical Physics, Stanford University, Stanford, California 94305, USA}
\affiliation{$^{2}$Department of Physics and Astronomy, University of California, Riverside, Riverside, CA 92511, USA}
\affiliation{$^3$SLAC National Accelerator Laboratory, 2575 Sand Hill Road, Menlo Park, CA 94025, USA}
\date{\today}

\begin{abstract}
Long wavelength descriptions of a half-filled lowest Landau level ($\nu = 1/2$) must be consistent with the experimental observation of particle-hole (PH) symmetry.  
The traditional description of the $\nu=1/2$ state pioneered by Halperin, Lee and Read (HLR) 
naively appears to break PH symmetry.
However, 
recent studies have shown that the HLR theory with weak quenched disorder can exhibit an emergent PH symmetry.  
We find that such inhomogeneous configurations of the $\nu=1/2$ fluid,  when described by HLR mean-field theory, are tuned to a topological phase transition between an integer quantum Hall state and an insulator of composite fermions with a dc Hall conductivity $\sigma_{xy}^{\rm (cf)} = - {1 \over 2} {e^2 \over h}$.
Our observations 
help explain why the HLR theory exhibits PH symmetric dc response.  
\end{abstract}

\maketitle

\section{Introduction}

The fractional quantum Hall effect, which is exhibited by two dimensional electron systems in a large perpendicular magnetic field, is a beautiful and mature subject.\cite{PrangeGirvin, dassarmaqhereview} 
A major reason for this success is that quantum Hall states are incompressible due to the presence of a gap to current carrying excitations. By contrast, the compressible states that occur near half-filled Landau levels have several aspects that remain unclear.  In this paper, we restrict our attention to the half-filled lowest Landau level (LLL) (filling fraction $\nu=1/2$).  

An important open question is whether the half-filled LLL with quenched disorder corresponds to a critical point or a stable phase of matter. When the disorder is sufficiently strong so that only integer quantum Hall plateaus exist, experiments reveal a direct transition  between a quantum Hall state, corresponding to filling fraction $\nu = 1$, and an insulator with $\nu = 0$ upon increasing the external magnetic field.\cite{Shahar1995, PhysRevLett.79.479}  
On the other hand, at weaker disorder, when some fractional quantum Hall states of the Jain sequence appear, experiments reveal that over a range of magnetic fields near $\nu =1/2$, the Hall resistivity $\rho_{xy}$ varies linearly with magnetic field, while the longitudinal resistivity $\rho_{xx}$ is non-zero.\cite{Willett97} 
This suggests that for weaker disorder, the half-filled Landau level is a stable compressible phase of matter.  
An understanding of how the physics of strong and weak disorder regimes evolve into one another remains elusive.

To address this question, one has to confront a notoriously difficult problem in which both disorder and strong interactions play an essential role. To make some progress on this issue, we employ the standard low-energy effective description of the LLL in terms of composite fermions.\cite{jain1989, lopezfradkin91, kalmeyerzhang, halperinleeread, simon1998, Jainbook, Fradkinbook}   
The nonperturbative effects of electron interactions in the LLL lead to a description in terms of composite fermions interacting with a fluctuating gauge field.   In the presence of quenched disorder,  it is conceivable that disorder effects may overpower those of gauge interactions among composite fermions.  
Within a mean-field treatment that ignores gauge fluctuations, we find that regardless of the disorder strength, the half-filled LLL always corresponds to a critical point rather than a stable phase of matter.  

What are the properties of such a critical point?  
A feature of two-dimensional electron systems is that the dc electrical resistivity tensor itself can be a universal amplitude, that aids in characterizing a critical point\cite{Fisher1990,Wen1990}.  We find that when the quenched disorder is self-averaging and preserves particle-hole symmetry of a half-filled lowest Landau level on average, the electrical Hall conductivity equals $\sigma_{xy} = e^2/2h$ at the critical point.  
This result holds over a broad range of disorder strengths and suggests the emergence of particle-hole symmetry.  

It should be stressed, however, that the particle-hole symmetry discussed here is a low energy property of a quantum critical point, and is not a microscopic symmetry.  
The latter possibility has been actively discussed as a property of electrons in a half-filled LLL interacting with two-body forces in the clean limit, and when Landau level mixing is absent (see the next section for a review).  
The manner in which such explicit particle-hole symmetry of an effective theory for a half-filled LLL is related to the emergent behavior at the critical point at $\nu=1/2$ is the main topic of the present paper.  

The paper is organized as follows.  In the next section, we review the standard description of the half-filled Landau level in terms of composite fermions pioneered by Halperin Lee and Read (HLR). 
We then focus on the issue of particle-hole symmetric response in the dc Hall conductivity of composite fermions. 
To this end, we use supersymmetric quantum mechanics to prove certain sum rules for the Hall conductivity of composite fermions which are corroborated by explicit numerical calculations. Then, with certain natural assumptions about the localization behavior of this theory, we show how the HLR theory of composite fermions describes an integer quantum Hall plateau transition.  In the final section, we discuss the low-energy equivalence between HLR theory at the integer quantum Hall transition critical point and theories with explicit particle-hole symmetry formulated in terms of Dirac composite fermions.

\section{Composite fermions and particle-hole symmetry: brief review}

Electrons in a half-filled lowest Landau level (LLL) are described by a $2+1$-dimensional Lagrangian of the form
\begin{equation}
 \mathcal L_{\rm el}= c^{\dagger} \left( i \partial_t + \mu +  A_t - \frac{1}{2 m_e} \left(i \partial_j + A_j \right)^2\right) c  + \ldots \nonumber
 \end{equation}
Here, $c(x)$ destroys a spin-polarized electron of mass $m_e$ at position $x = (t, \bm r)$, $A_t({\bm r})$ is a (static) electromagnetic scalar potential, $\bm A(\bm r) = (A_x(\bm r), A_y(\bm r))$ is the electromagnetic vector potential corresponding to the uniform perpendicular magnetic field $B = \nabla \times \bm A(\bm r) > 0$, and $\mu$ is the chemical potential adjusted such that the Landau level is half-filled, $ B = 4 \pi \langle c^{\dagger} c\rangle$.\footnote{In the remainder, we choose units such that $e^2 = \hbar = 1$. Thus, the quantum of conductance $e^2/h = 1/2\pi$.}
The simplicity of the above Lagrangian is deceptive: 
the effects of the interactions denoted by $``\ldots"$ are singular due to the extensive degeneracy of a partially filled Landau level in the clean limit.  

A conceptually simpler approach involves emergent particles known as composite fermions.\cite{jain1989, lopezfradkin91, kalmeyerzhang, halperinleeread, simon1998, Jainbook, Fradkinbook}
In the composite fermion theory of Halperin, Lee and Read (HLR),\cite{halperinleeread, kalmeyerzhang} the low-energy behavior at $\nu=1/2$ is postulated to be governed by the Lagrangian:
 \begin{align}
 \mathcal L_{\rm HLR} = \mathcal L_f + \mathcal L_{\rm cs} + \ldots, \nonumber 
 \end{align}
where
\begin{align}
\mathcal L_f  & = f^{\dagger} \left( i \partial_t  + \mu +  A_t + a_t - \frac{1}{2 m} \left(i \partial_j + A_j + a_j \right)^2 \right) f, \cr
\mathcal L_{\rm cs} & = \frac{1}{2} \frac{1}{4 \pi} a d a, \nonumber
\end{align}
$f(x)$ destroys a composite fermion of mass $m$,  $a_{\alpha}(x)$ for $\alpha \in \{t,x,y\}$ is an emergent $U(1)$ gauge field with a Chern-Simons term $a d a \equiv \epsilon^{\alpha \beta \gamma} a_{\alpha} \partial_{\beta} a_{\gamma}$ that implements flux attachment, where the anti-symmetric tensor $\epsilon^{txy} = 1$.\footnote{Throughout, we consider the simplified, albeit less precise, versions of various composite fermion Lagrangians in which certain Chern-Simons terms have non-quantized levels that do not produce gauge-invariant terms in an effective action. 
For instance, ${\cal L}_{\rm cs}$ should be viewed as short-hand for $- {1 \over 2\pi} \epsilon^{\mu \nu \lambda} a_{\mu} \partial_{\nu} b_{\lambda} - {2 \over 4 \pi} \epsilon^{\mu \nu \lambda} b_{\mu} \partial_{\nu} b_{\lambda}$,\cite{Fradkinbook} while $- \frac{1}{4 \pi} \epsilon^{\mu \nu \lambda} a_{\mu} \partial_{\nu} A_{\lambda} + \frac{1}{8\pi} \epsilon^{\mu \nu \lambda} A_{\mu} \partial_{\nu} A_{\lambda}$ in ${\cal L}_{\rm Dirac}$ below should be understood as $- \frac{1}{8 \pi} \epsilon^{\mu \nu \lambda} a_{\mu} \partial_{\nu} a_{\lambda} - \frac{2}{4\pi} \epsilon^{\mu \nu \lambda} b_{\mu} \partial_{\nu} b_{\lambda} + {1 \over 2\pi} \epsilon^{\mu \nu \lambda} b_{\mu} \partial_{\nu} (a_{\lambda} - A_\lambda)$.\cite{Seiberg:2016gmd} } 
By virtue of the $a_t$ equation of motion, 
\begin{align}
4 \pi f^\dagger f(\bm r) = - \nabla \times \bm a(\bm r),
\end{align}
two units of flux are attached to each composite fermion.  
Since this flux attachment does not change the density of particles, $f^\dagger f(\bm r) = c^\dagger c (\bm r)$, at half-filling the composite fermions feel on average a vanishing effective magnetic field $B_{\rm eff} = \nabla \times (\bm A + \bm a)$.
In a mean-field approximation, the fluctuations of $a_\alpha$ are neglected.
The resulting mean-field ground state is a system of finite density fermions in zero magnetic field: {\it i.e.}, a filled Fermi sea of composite fermions.
Composite fermion mean-field theory naturally accounts for the fact that the $\nu=1/2$ state is compressible;
it has led to many other successful experimental predictions including that of large cyclotron orbits dictated by the much smaller effective field $B_{\rm eff}$, rather than the applied field $B$ in the vicinity of half-filling.\cite{Willett97}

However, low-energy effective theories must be consistent with observed symmetries.   
An important constraint comes from 
particle-hole (PH) symmetry,\cite{PhysRevLett.50.1219, girvin1984}
which is an exact symmetry of a half-filled lowest Landau level of electrons in the limit of zero Landau level mixing and of an infinitely strong magnetic field.
Under a particle-hole transformation, electrons filling an empty Landau level transform to holes depleting a filled Landau level.  Under this operation, the zero-temperature dc electrical Hall conductivity\cite{girvin1984,kivelson1997}
\begin{equation}
\sigma_{xy} \rightarrow \frac{1}{2 \pi} - \sigma_{xy}.
\end{equation}
Since the half-filled Landau level is equivalently described either by electrons or holes, particle-hole symmetry requires $\sigma_{xy} = 1/4 \pi$.  
In fact, PH symmetry requires this for all frequencies below some characteristic scale.
In this paper, however, we are only concerned with the zero-temperature dc response.
Remarkably, particle-hole symmetric electrical response is found experimentally at $\nu=1/2$.\cite{Shahar1995, Wong1996}

What are the implications of particle-hole symmetry for composite fermions?  
It turns out that in order that the electrical Hall conductivity $\sigma_{xy} = 1/4\pi$, the composite fermion Hall conductivity in HLR theory, defined in terms of linear response to $a_{\mu}$ must satisfy\cite{kivelson1997}
\begin{equation}
\label{phcf}
\sigma_{xy}^{(\rm cf)} = -\frac{1}{4 \pi},
\end{equation}
whenever the electrical resistivity $\rho_{xx} \neq 0$.
This may be viewed as a necessary condtition for satisfying 
particle-hole symmetry at $\nu=1/2$.  
Naively, however, this relation seems to be badly violated in HLR theory, since the composite fermions see on average a vanishingly small effective magnetic field.
Why would such a system have a Hall conductivity on the order of the quantum of conductance?  
For this reason, it was suspected that the HLR theory failed to satisfy particle-hole symmetry.\cite{kivelson1997, BMF2015}

Recently, an alternative theory, introduced by Son,\cite{Son2015} conjectured that a particle-hole symmetric half-filled Landau level is described by Dirac composite fermions $\psi$ ($\bar{\psi} = \psi^\dagger \gamma^0$)
with Lagrangian:
\begin{equation}
\label{Son}
\mathcal L_{\rm Dirac} = i \bar \psi \gamma_{\nu} D_a^{\nu} \psi - \frac{1}{4 \pi} A d a + \frac{1}{8\pi} A d A,
\end{equation}
where $D^\nu_a = \partial_\nu - i a_\nu$, $a_\nu$ is again an emergent $U(1)$ gauge field, and $\gamma_\nu$ are appropriate $2 \times 2$ $\gamma$-matrices.
The Dirac composite fermions above only carry emergent gauge charge and are electromagnetically neutral.  
The Dirac theory is manifestly particle-hole symmetric.   At first sight, the Dirac and HLR theories seem drastically different from one another, 
and while the Dirac theory preserves PH symmetry, the HLR theory appears to violate it.

Surprisingly, several recent papers,\cite{PhysRevB.95.235424, 2017PhRvX...7c1029W, 2018arXiv180307767K} 
have found that HLR theory {\it can} exhibit PH symmetric response
provided that inhomogeneous configurations (be it a spatially varying external potential or quenched disorder) explicitly be taken into account.
(See Refs.~\onlinecite{PotterSerbynVishwanath2015, PhysRevB.94.245107, BalramJain2016, LevinSon2016} for other studies contrasting the HLR and Dirac composite fermion theories and earlier numerical work in Ref.~\onlinecite{rezayi2000} that found large overlap between the wave functions corresponding to the HLR composite fermion theory and ground states of particle-hole symmetric electron Hamiltonians at $\nu=1/2$.)
Our goal in the remainder of this paper is to explain this behavior, {\it i.e.}, why HLR theory can exhibit particle-hole symmetric response: we will show that HLR mean-field theory at $\nu = 1/2$ describes a critical point between an integer quantum Hall state ($\sigma_{xy}^{\rm (cf)} = - 1$) and an insulator ($\sigma_{xy}^{\rm (cf)} = 0$) of composite fermions with particle-hole symmetry dc Hall response (Eq.~\eqref{phcf}).


\section{HLR theory with weak quenched disorder }

We first rewrite the HLR Lagrangian in terms of a shifted gauge field $a_\mu \rightarrow a_\mu - A_\mu$: 
 \begin{align}
 \label{shifted}
\mathcal L_{\rm HLR} = \mathcal L_f + \mathcal L_{\rm cs} + \ldots
\end{align}
with
\begin{align}
\label{compositefermion}
\mathcal L_f  & = f^{\dagger} \left( i \partial_t  + \mu + a_t - \frac{1}{2 m} \left(i \partial_j + a_j \right)^2 \right) f \\
\label{csterm}
\mathcal L_{\rm cs} & = \frac{1}{2} \frac{1}{4 \pi}  \left(a - A \right) d  \left(a - A\right).
\end{align}
We will  consider spatially inhomogeneous configurations of composite fermions.  
We do so by adding a weak spatially-varying correction to the chemical potential:
\begin{equation}
\mu \rightarrow \mu(\bm r) = \mu + V(\bm r), \ \ |V(\bm r)| \ll \mu.
\end{equation}
The quantity $V(\bm r) $ is a quenched random variable with 
\begin{align}
\label{quencheddis}
\overline{V(\bm r)} = 0, \quad \overline{V(\bm r) V(\bm r')} = \sigma_V^2 e^{- |\bm r - \bm r'|^2/{\cal R}^2},
\end{align}
where the overline denotes averaging with respect to disorder.

In linear response to $V(\bm r)$, there will be a spatial variation of the composite fermion density:
\begin{align}
 \langle f^{\dagger} f(\bm r) \rangle = n({\bm r}) = \chi V({\bm r}) + \overline{n},
 \end{align}
where $\chi$ is the composite fermion static compressibility and $\bar n \equiv B/4 \pi$ is the average density at $\nu=1/2$.  
Neglecting gauge fluctuations, the equation of motion of $a_t$  sets the relation between the composite fermion density and flux variations:
\begin{align}
n(\bm r) & = \frac{B - b(\bm r)}{4 \pi} \cr
& =  \bar n - \frac{b(\bm r)}{4 \pi},
\end{align}
where $b(\bm r) = \epsilon_{ij} \partial_i a_j(\bm r)$.
Therefore, $V(\bm r)$ gives rise to the spatially-varying magnetic flux $b(\bm r)$:
\begin{equation}
\label{slaving}
b(\bm r)  = -4 \pi \chi V(\bm r) = - 2 m V(\bm r),
\end{equation}
whose statistical properties are determined by Eq.~\eqref{quencheddis}.
In Eq.~\eqref{slaving}, we identified $\chi = m/2\pi$ with the long wavelength static compressibility of a free fermion gas. Since the wavevector dependent compressibility $\chi(\bm q) =\chi(\bm 0)= m/2\pi$ for $|\bm{q}|<2k_F$, we can expect this assumption to hold locally if the disorder is sufficiently long ranged so that the Fourier components of disorder for $|\bm{q}| > 2k_F$ can be taken to be zero.
Recent work\cite{PhysRevB.95.235424, 2017PhRvX...7c1029W, 2018arXiv180307767K} (see Ref.~\onlinecite{kivelson1997} for an earlier discussion) has found that Eq.~\eqref{slaving}, valid in the weak disorder limit when gauge fluctuations are negligible, is an essential ingredient for effecting particle-hole symmetric behavior of the HLR composite fermion theory.
  
Incorporating the constraint Eq.~\eqref{slaving} that determines $a_j$ in terms of $V(\bm r)$ into $\mathcal L_f$, the composite fermion Lagrangian becomes
\begin{align}
\label{cflagmod}
\mathcal L_f = f^{\dagger} \left( i \partial_t  +\mu + a_t  - \frac{b(r)}{2m} - \frac{1}{2 m} \left(i \partial_j + a_j \right)^2 \right) f.
\end{align}
At the mean-field level where $a_\mu$ is taken to be a static background field, the dynamics of composite fermions is governed entirely by the Lagrangian ${\cal L}_f$ in Eq.~\eqref{cflagmod} or, equivalently, the free composite fermion Hamiltonian:
\begin{equation}
\label{meanfieldH}
H_f = \frac{1}{2 m} \left[ \Pi_x^2 + \Pi_y^2 + b(\bm r) \right] , \ \ \Pi_{j} = - i \partial_j - a_j.
\end{equation}
This Hamiltonian describes a Landau level problem of a system of noninteracting composite fermions with gyromagnetic ratio $g=2$. 
Note, however, that only one spin species is present.
It is important to remember that $g$ is determined by the composite fermion compressibility $\chi$.



\subsection{Intuitive argument\label{intuitiveargument}}

In order to gain some intuition on the effects of disorder on the free nonrelativistic composite fermion, consider two equivalent, representative regions of space $R_1$ and $R_2$: the first with $V(\bm r) \approx - V_0 < 0$ for $\bm r \in R_1$ and the second with $V(\bm r) \approx V_0 > 0$ for $\bm r \in R_2$.    
With long wavelength disorder (${\cal R} k_F \gg 1$), the characteristic sizes of such  regions will be large compared to the Fermi wavelength.  
The local Hamiltonians governing the two bulk regions are
\begin{align}
H_1 =& \frac{1}{2m} \left[ (\bm p + \bm a)^2 + |b_0| \right], 
\\
H_2 =& \frac{1}{2m} \left[ (\bm p - \bm a)^2 - |b_0| \right], 
\end{align}
where $\nabla \times \bm a = b_0 = - 2m V_0 < 0$. 
\footnote{The vector potentials in $H_1$ and $H_2$ are not simply related by a change of sign in general. They will have an extra gauge transformation piece that is needed to ensure continuity of $\bm a$ in space.}
The (approximate) bulk spectrum in each region is trivially obtained:
\begin{align}
\label{spectrum1}
E_1 & = \frac{|b_0 |}{m} (n+1) , \\
\label{spectrum2}
E_2 & = \frac{|b_0 |}{m} n, 
\end{align}
where $n$ is a non-negative integer.  
Since $b_0 \ll \mu$, we expect that many Landau levels will be filled in both regions.   
However, there is an ``unpaired" Landau level at zero energy in the $V > 0$ region, whereas the finite energy spectra are identical in both regions.
 
This mismatch of spectra in the two regions due to the zero mode  has an immediate physical consequence: there will be one extra Landau level filled in region $R_2$.  
If we suppose that the localization length due to disorder is small compared to the range $\mathcal R$ of the disorder, then we may write an effective response Lagrangian governing the two regions,
 \begin{equation}
 \label{effLagrangian1}
 \mathcal L_{\rm response} = \frac{N}{4 \pi} a d a - \frac{N+1}{4 \pi} a d a,
 \end{equation}
where $N$ is a large positive integer corresponding to the number of filled Landau levels in the $V < 0$ region and the two Chern-Simons terms have support on regions $R_1$ and $R_2$.  
Upon volume averaging all such regions, we find the following contribution to the averaged Hall conductivity, valid on lengths scales large compared the size of any individual region:
 \begin{equation}
\sigma^{\rm avg}_{xy} = - \frac{1}{4 \pi}. \label{cfhallconductivity}
 \end{equation}

\subsection{Supersymmetric quantum mechanics\label{susyQM}}

The spectra in Eqs.~\eqref{spectrum1} and \eqref{spectrum2} are identical to that of a two-dimensional spin-1/2 particle of charge $e=1$ in a transverse magnetic field $b_0 < 0$ with Hamiltonian:
\begin{equation}
{\cal H} = \frac{ \Pi_x^2 + \Pi_y^2}{2m} \mathbb{I}_{2 \times 2} - \frac{g}{4 m} b_0  \sigma_z
\end{equation}
at $g=2$.
Here,  $\mathbb{I}_{2 \times 2}$ is a $2\times 2$ identity matrix, $\sigma^z$ is the usual Pauli-matrix, and ``spin-up" (``spin-down") components are labeled by their $\sigma_z$ eigenvalue $1 (-1)$ and correspond to the $V < 0$ ($V > 0$) regions. 
Although the spectra of $H_1$ and ${\cal H}_\uparrow$ defined above are same, the Hamiltonians are different. 
They differ from each other by a time-reversal operation. 
If we define $\mathcal{T}$ to represent a time-reversal transformation, which changes the sign of the vector potential while leaving the sign of the scalar potential invariant, under which the HLR Lagrangian is not invariant, we find $ \mathcal{H}_\uparrow = \mathcal{T}^{-1} H_1 \mathcal{T}$.
\footnote{Our definition of the time-reversal transformation may seem unnatural from the perspective of supersymmetric quantum mechanics where only a vector potential appears in the spin-up and spin-down Hamiltonians when written as $QQ^\dagger$ and $Q^\dagger Q$. It faithfully represents the fact that the potential terms, that arise upon expanding $QQ^\dagger$ and $Q^\dagger Q$ in terms of $\Pi^2_j$ and $b(\bm r)$, are even under the time-reversal transformation defined through HLR Lagrangian; otherwise, the individual spin-up and spin-down Hamiltonians would be statistically time-reversal invariant, in contradiction with the HLR composite fermion and composite hole Lagrangians.}
On the other hand, $\mathcal{H}_\downarrow = H_2 $.


Now consider spatially-varying potential and flux with the spin-up and spin-down components defined on all of space.
The spin-up and spin-down components of the Hamiltonian ${\cal H} = \cal H_\uparrow \oplus \cal H_\downarrow$  become
\begin{align}
\label{spinuphamiltonian}
{\cal H}_\uparrow & = {1 \over 2m} \Big[\Pi_x^2 + \Pi_y^2 - {g \over 2} b(\bm r) \Big], \\
\label{spindownhamiltonian}
{\cal H}_\downarrow & = {1 \over 2m} \Big[\Pi_x^2 + \Pi_y^2 + {g \over 2} b(\bm r) \Big],
\end{align}
with $g=2$.
The spin-down Hamiltonian ${\cal H}_\downarrow$ is the HLR mean-field Hamiltonian $H_f$ in Eq.~\eqref{meanfieldH}. 
Although the spin-up component is not present per se in the physical problem of interest, we will show that it is useful to introduce it in order to prove that the composite fermions satisfy Eq.~\eqref{cfhallconductivity}.
One can however think of $\mathcal{H}_\uparrow$ as being naturally present in an ensemble of disorder realizations. Since the disorder is particle-hole symmetric, for every disorder realization $V(\bm r)$ there is another disorder realization $-V(\bm r)$ in the ensemble. The spin-up component can be considered to be the time-reversed Hamiltonian of the latter disorder configuration.
Consequently, the spin-up particles experience a slaved disorder satisfying $b(\bm r) = 2 m V(\bm r)$, rather than the constraint in Eq.~\eqref{slaving} that relates the potential and flux disorder felt by the spin-down particles.

The utility of defining $\mathcal{H}_\uparrow$ and $\mathcal{H}_\downarrow$ this way is that ${\cal H}$ realizes supersymmetric quantum mechanics.
To see this, we define the operators:
\begin{eqnarray}
Q &=& \frac{\Pi_x - i\Pi_y}{\sqrt{2m}}, \\
Q^\dagger &=&   \frac{\Pi_x + i\Pi_y}{\sqrt{2m}}.
\end{eqnarray}
Notice that these generalized Landau level ladder operators remain well defined for arbitrary spatially-varying magnetic field. 
Since $[\Pi_x, \Pi_y] = i b(\bm{x})$, we find $[Q, Q^\dagger] = - b(\bm r)/m$. 
At $g=2$ only, we can write the Hamiltonians for the two spin components in a supersymmetric form:
\begin{eqnarray}
\label{susyone}
\mathcal{H}_\uparrow &=& Q Q^\dagger,\\
\label{susytwo}
\mathcal{H}_\downarrow &=& Q^\dagger Q.
\end{eqnarray}
As expected of a supersymmetric spectrum, the eigenvalues of the two Hamiltonians are non-negative. 

More importantly, supersymmetry guarantees that there is an exact mapping between the finite-energy eigenvalues and eigenstates of the spin-up and spin-down Hamiltonians.
If $\ket{\psi_{\downarrow}}$ is a normalized eigenstate of $\mathcal{H}_\downarrow$ with energy $E  > 0$, then $\ket{\psi_{\uparrow}} = \frac{Q}{\sqrt{E}}\ket{\psi_\downarrow}$ is a normalized eigenstate of $\mathcal{H}_\uparrow$ with the same energy. 
This spectral equivalence implies that if eigenstates at a particular non-zero energy are localized (extended) for $\mathcal{H}_\uparrow$, then they will be localized (extended) for $\mathcal{H}_\downarrow$ as well.
Supersymmetry will enable us to make analytical arguments -- that we will supplement with a numerical calculation -- for the Hall conductivity of the spin-down or HLR composite fermions.  

\subsection{Hall conductivity sum rules}

We now argue that the dc Hall conductivities of the spin-down and spin-up eigenstates of ${\cal H}$ satisfy the following two sum rules in the presence of weak disorder:
\begin{align}
\label{sumfirst}
\overline{\sigma}_{xy}^\downarrow + \overline{\sigma}_{xy}^\uparrow & = 0, \\
\label{sumsecond}
\sigma_{xy}^\downarrow - \sigma_{xy}^\uparrow & = - {1 \over 2 \pi}.
\end{align}
The first equation is valid for disorder averaged Hall conductivities (denoted by an overline), while the second equations is valid for each disorder realization. These sum rules say that the spin-1/2 system defined by ${\cal H}$ has a vanishing electrical Hall conductivity (Eq.~\eqref{sumfirst}) and exhibits an integer spin Hall response (Eq.~\eqref{sumsecond}).
A consequence of these sum rules is clear: $\sigma_{xy}^{\rm (cf)} \equiv \overline{\sigma}_{xy}^\downarrow = - 1/4\pi$. 

\subsubsection{First sum rule}

To establish the first sum rule in Eq.~\eqref{sumfirst}, it is sufficient to show that ${\cal H}$ has a statistical time-reversal symmetry. 
Since both spin-up and spin-down components are present and the magnetic field is zero on average (and all of its higher odd moments vanish as well), we expect this to be true. 

More formally, we can define the time-reversal operation in the spin-space, $\mathcal{T}_s = -i\sigma_y \mathcal{T}$. Under this time-reversal transformation, the Hamiltonian,
\begin{align}
\mathcal{H} &\mapsto \mathcal{T}_s^{-1} \mathcal{H} \mathcal{T}_s \nn \\
& =  \begin{pmatrix}
\mathcal{T}^{-1} \mathcal{H}_\downarrow\mathcal{T} & 0 \\
0 & \mathcal{T}^{-1} \mathcal{H}_\uparrow\mathcal{T}
\end{pmatrix} \nn\\
& = \frac{1}{2m}\begin{pmatrix}
(\bm p+\bm a)^2 + b(\bm r) & 0 \\
0 & (\bm p+\bm a)^2 - b(\bm r)
\end{pmatrix}.
\end{align}
The vector potential $\bm a \rightarrow -\bm a$, the scalar potential (represented by the $b(\bm r)$ term) is invariant, and spin-up and spin-down are interchanged. 
This means that $\sigma_{xy}^\uparrow \rightarrow -\sigma_{xy}^\downarrow$ and $\sigma_{xy}^\downarrow \rightarrow -\sigma_{xy}^\uparrow$ so that the total Hall condutivity $\sigma_{xy}^\downarrow + \sigma_{xy}^\uparrow$ changes sign. 
From Eqs.~\eqref{spinuphamiltonian} and~\eqref{spindownhamiltonian}, this is equivalent to changing the sign of the potential.
If the disorder ensemble is particle-hole symmetric, $V(\bm r)$ and its particle-hole conjugate $-V(\bm r)$ will appear with equal weights.
Consequently, the sum of the spin-up and spin-down Hall conductivity
is equal in magnitude but opposite in sign for these two disorder configurations and its disorder average:
%
\begin{eqnarray}
\overline{\sigma}_{xy}^\downarrow + \overline{\sigma}_{xy}^\uparrow & = 0.
\end{eqnarray}

\subsubsection{Second sum rule}

Establishing the second sum rule in Eq.~\eqref{sumsecond} is more involved.
In fact, the presence of weak disorder is generally necessary for it to be satisfied.
Our strategy is to first consider the spin Hall conductivity when
\begin{align}
\label{constantfluxpart}
{1 \over L^2} \int d^2 {\bm r}\ b(\bm r) = b_0 \neq 0
\end{align}
where $L^2$ is the area of our system.
Physically, $b_0$ corresponds to the effective magnetic field experienced by the composite fermions.
By studying the spin Hall conductivity for non-zero $b_0$, we show analytically that $b_0 = 0$ is a transition point at which both the spin-up and spin-down Hall conductivities change by $-1/2\pi$.
We then confirm numerically that $\overline{\sigma}_{xy}^\downarrow = - \overline{\sigma}_{xy}^\uparrow = - 1/4\pi$ at $b_0 = 0$.

\begin{figure}[ht]
\includegraphics[width=2.5in]{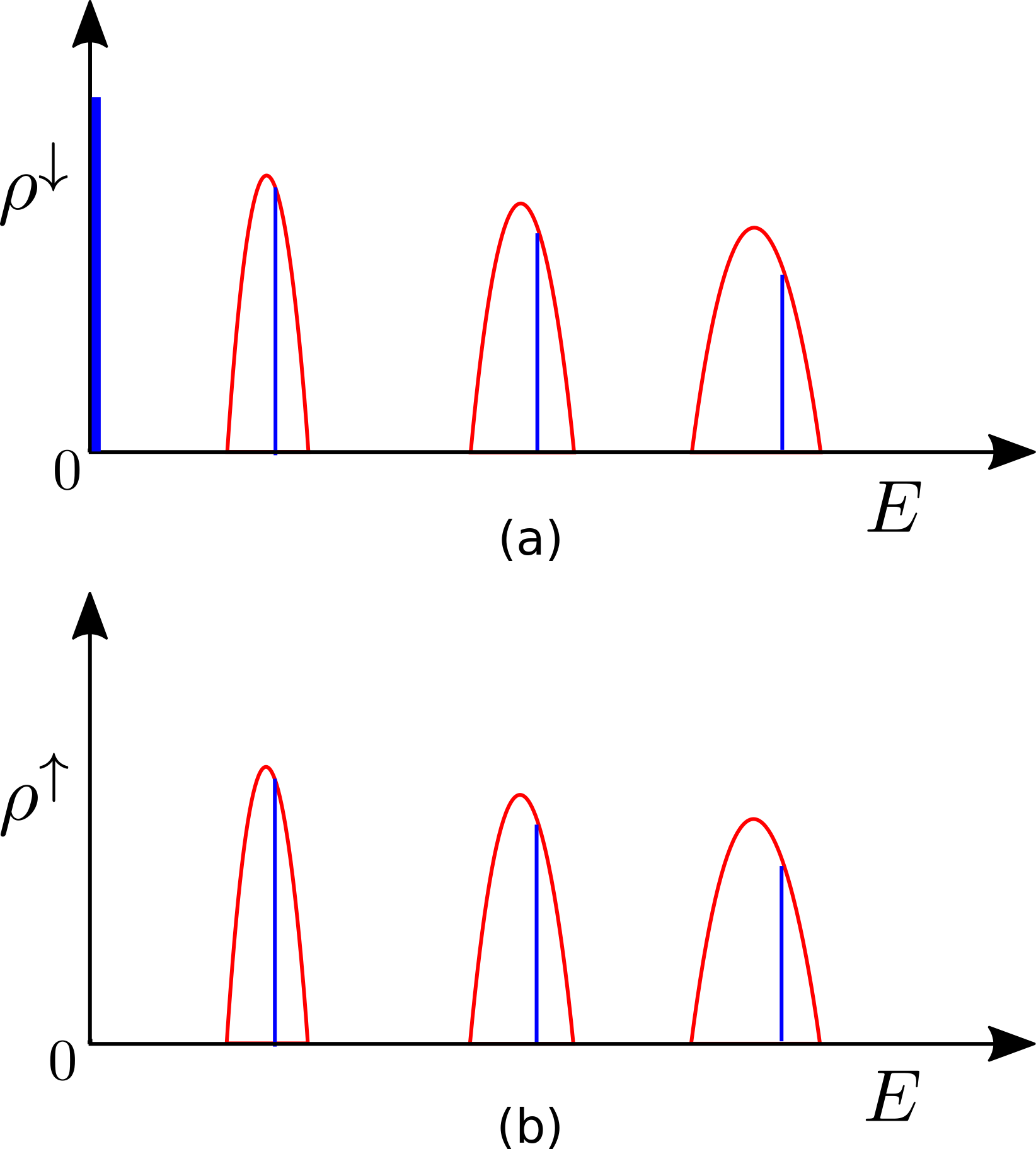}
\caption{A schematic for the density of states of (a) down-spin and (b) up-spin when average magnetic field $b_0 <0$ and spatial variations of $b(\bm r)$ are small compared to $|b_0|$. The lowest Landau level of down-spins at zero-energy does not spread as long as $g=2$. Also, according to Eq.~\eqref{LL_energy_upspin} and~\eqref{LL_energy_downspin}, the higher Landau levels spread more in energy because to a leading order $E_n^\downarrow\propto nb(x)$. All states except one are assumed to be localized in each positive energy Landau level. As a result of supersymmetric quantum mechanics, the locations of extended and localized states match for the two spins. Anticipating levitation\cite{PhysRevLett.52.2304} of extended states as the spatial variations of magnetic field are made larger, we have drawn the extended state at a higher energy than the center of each Landau level.}
\label{DoS_weak_variations}
\end{figure}

We begin by explaining why we expect this sum rule to hold. 
Write $b(\bm r) = b_0 + \tilde{b}(\bm r)$, where $\tilde{b}(\bm r)$ has zero average and corresponds to the spatially varying part of the magnetic field. For definiteness, consider $b_0 < 0$. When $\tilde{b}(\bm r) = 0$, one has sharp Landau levels at energies given by:
\begin{eqnarray}
E^\uparrow_{n} &=&  \frac{|b_0|}{m} (n+1) \label{LL_energy_upspin}\\
E^\downarrow_{n} &=&  \frac{|b_0|}{m}n \label{LL_energy_downspin}.
\end{eqnarray}
At a given Fermi energy, the spin-downs have exactly one more Landau level filled than the spin-ups. 
Since, $b_0 < 0$, each Landau level contributes $\sigma_{xy} = -\frac{1}{2\pi}$ to the Hall conductance.
The additional filled Landau level implies that the difference of the spin-up and spin-down Hall conductivities satisfies Eq.~\eqref{sumsecond}. Now add weak spatial variations so that $\tilde{b}(\bm r) \neq 0$ and $|\tilde{b}(\bm r)| \ll |b_0|$. 
A schematic of the density of states is shown in Fig.~\ref{DoS_weak_variations}. 
Since our system consists of non-interacting fermions, for all positive energy Landau levels, all states are presumably localized except at one energy, where the states are extended. 
It should be noted that the zero-energy Landau level of spin-downs does not spread in energy. 
This is so because the spin-ups cannot have states close to the zero energy if the spatial variations of magnetic field are small. 
Supersymmetric quantum mechanics implies that the localized states and extended states lie at exactly the same energies in the finite-energy spectra of the spin-up and spin-down Hamiltonains. 
Thus, we continue to expect that Eq.~\eqref{sumsecond} remains true.
%
Identical considerations hold when $b_0>0$.
In this case, the plots of density of states of spin-downs and spin-ups are reversed so that spin-ups will contribute one extra quantum of Hall conductance. 
However, since $b_0 > 0$, the spin-up and spin-down Hall conductivities will be positive.
Thus, the sum rule in Eq.~\eqref{sumsecond} is satisfied for any sign of $b_0$ as long as spatial variations of magnetic field are weak compared to $|b_0|$.


Now, we would like to show that Eq.~\eqref{sumsecond} is satisfied for arbitrary $b(\bm r)$. 
In this regard, supersymmetry plays a crucial role.
Since supersymmetry provides a one-to-one map between finite-energy spin-up and spin-down eigenstates, we expect such paired states to make identical contributions to their respective Hall conductivities.
In Appendix \ref{Efindependence}, we use supersymmetric quantum mechanics and the Kubo formula to show that 
%
\begin{align}
\label{fermiindependence}
{\partial \over \partial \mu} \Big( \sigma_{xy}^\downarrow -\sigma_{xy}^\uparrow \Big) = 0
\end{align}
whenever $b_0 \neq 0$ and the Fermi energy $\mu > 0$.
In proving Eq.~\eqref{fermiindependence}, we did not need to make specific assumptions about the strength of the magnetic field variations or whether the finite-energy states are localized or extended. 
However, Eq.~\eqref{fermiindependence} does rely on unbroken supersymmetry.

It remains to determine the contribution of any states at zero energy to $\sigma_{xy}^\downarrow - \sigma_{xy}^\uparrow$.
The zero-energy states of $\mathcal{H}_\uparrow$ in Eq.~\eqref{susyone} are annihilated by $Q^\dagger$ and those of $\mathcal{H}_\downarrow$ are annihilated by $Q$.
Consequently, there is no correspondence between possible zero-energy states of ${\cal H}_\downarrow$ and ${\cal H}_\uparrow$.
For any non-zero $b_0$, there is a spectral mismatch at zero energy.
While both ${\cal H}_\uparrow$ and ${\cal H}_\downarrow$ may admit zero-energy eigenstates when $b_0 \neq 0$, their normalizability depends on the sign of $b_0$.
When $b_0 < 0$, only ${\cal H}_\downarrow$ has normalizable zero-energy states (analogous to the spectra in Eqs.~\eqref{spectrum1} and \eqref{spectrum2}) and vice versa when $b_0 > 0$.\cite{PhysRevA.19.2461}
We would like to argue that these zero-energy states contribute $1/2\pi$ ($-1/2\pi$) to the up-spin (down-spin) Hall conductivity for positive (negative) $b_0$ regardless of the strength of $b(\bm r)$.

To see this, let's take $b_0 < 0$.
(Identical considerations apply when $b_0 > 0$.)
Reminiscent of a Landau level, the number of normalizable zero-energy states of ${\cal H}_\downarrow$ is equal to the number of flux quanta $N_\Phi = |b_0| L^2/2\pi$ passing through the $L \times L$ planar system.\cite{PhysRevA.19.2461, PhysRevD.29.2375}
Importantly, this is true for arbitrary magnetic fields $b(\bm r)$ as long as Eq.~\eqref{constantfluxpart} is satisfied. 
This is also consistent with the earlier observation that the Landau level at zero energy does not spread in energy as $|\tilde b(\bm r)|$ is increased in strength.

We can now use Laughlin's flux threading argument.\cite{PhysRevB.23.5632} 
Because spin-ups do not have any normalizable zero-energy modes and there is a one-to-one correspondence between the positive energy eigenstates of spin-downs and spin-up, under a threading of flux, zero-energy states cannot continuously lift up in energy. 
After exactly one flux quantum is adiabatically threaded, a filled set of zero-energy states transforms back to itself. 
Thus, the only change that could have happened is that an integer number of spin-down fermions got transferred across the system. Consequently, the Hall conductivity of a filled set of zero-energy states is quantized. 
We can expect that upon adiabatic switching on of the spatial variations $\tilde b (\bm r)$, the Hall conductivity should not change continuously. 
Thus, the zero-energy states contribute $-1/2\pi$ to $\sigma_{xy}^\downarrow - \sigma_{xy}^\uparrow$. Importantly, this is true even when the spatial variations $\tilde b(\bm r)$ are large compared to $|b_0|$ and the gap between zero- and positive-energy states vanishes.
%

To better understand the topological aspect of zero-energy states and their contribution to the spin-down Hall conductivity for $b_0 < 0$, it is useful to consider the special case when disorder is a function of a single spatial coordinate $x$.
We choose the gauge where the vector potential ${\bm a} = (\tilde{a}_y(x) + b_0 x) \hat{\bm y}$ with $\partial_x \tilde{a}_y(x) = \tilde{b}(x)$.
Zero-energy states satisfy the equation, $Q \Psi = 0$, i.e.,
\begin{align}
\label{zeroenergyequation}
\Big(\partial_x - i \partial_y - a_y\Big) \Psi(\bm r) = 0.
\end{align}
Introducing a scalar function $\Phi(\bm r) = \tilde{\Phi}(x) - b_0 {x^2 \over 2}$ that satisfies $\partial_x \Phi = -a_y $ so that $\nabla^2 \Phi = - b(x)$, the solutions to Eq.~\eqref{zeroenergyequation} take the form:\cite{PhysRevD.29.2375}
\begin{align}
\Psi_k(\bm r) = {\cal N} e^{i k y} e^{{b_0 \over 2} (x - x_k)^2} e^{-\tilde{\Phi}(x)}
\label{cylindrical_zero_modes},
\end{align}
where $k$ is the momentum carried by the state in the $y$-direction, $x_k = k/b_0$, and ${\cal N}$ is a normalization constant.
The zero-energy wave functions $\Psi_k$ are normalizable since $b_0 < 0$.
(Zero-energy states of ${\cal H}_\uparrow$ have the same form with the replacements $y \mapsto - y$, $b_0 \mapsto - b_0$ and $\tilde \Phi(x) \mapsto -\tilde\Phi(x)$.)
When the fluctuating component of $b(\bm r)$ is zero, $\tilde{\Phi}(x) = 0$ and the states $\Psi_k$ precisely coincide with those of a lowest Landau level in a uniform magnetic field, as expected.

We again use Laughlin's flux-threading argument to determine the contribution of the $\Psi_k$ to the down-spin Hall conductivity.
To this end, we take the planar system to be a $L \times L$ torus so that $k \sim k+ |b_0 L|$ becomes quantized in units of $2\pi/L$ and $\tilde{\Phi}(x) = \tilde{\Phi}(x + L)$.
Direct calculation shows that
\begin{align}
\label{xvevs}
\langle \Psi_{k+|b_0| L} | \hat{x} | \Psi_{k+|b_0| L} \rangle - \langle \Psi_k | \hat{x} | \Psi_k \rangle =- L.
\end{align}
 \begin{figure}[ht]
\includegraphics[width=2.7in]{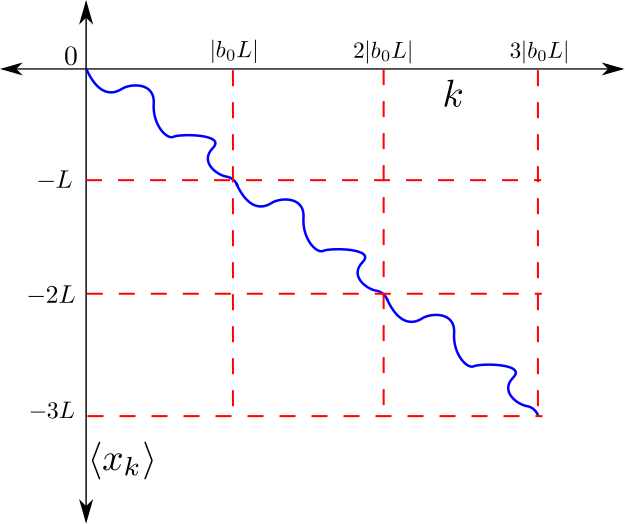}
\caption{A schematic of $\langle x_k \rangle$ vs $k$ for zero-energy wavefunctions in Eq.~\eqref{cylindrical_zero_modes}. As seen in Eq.~\eqref{xvevs}, $x_k$ is a periodic function of $k$. Or, if $k \rightarrow k + |b_0|L$, $x_k \rightarrow x_k - L$. Let's assume that $k=0$ through $k = |b_0| L - 2\pi/L$ are filled initially. After adiabatic threading of one flux quantum, $k \rightarrow k+2\pi/L$. Thus, $k=2\pi/L$ through $k=|b_0|L$ are filled now. Equivalently, $k=0$ mode is now replaced by $k = |b_0|L$ and rest of the modes are unaffected. Since, the separation between these two modes is $-L$, one down-spin fermion got transported across the system in negative x-direction.}
\label{xk_cylindrical}
\end{figure}
States that differ by $N_\Phi$ units of momentum are separated by a distance $L$ around the $x$-direction.
As we thread $2\pi$ flux by adiabatically varying $a_y \rightarrow a_y + 2\pi/L$, the zero-energy states evolve back into themselves upon shifting $k \rightarrow k + 2\pi/L$.
In this process, Eq.~\eqref{xvevs} implies that a single spin-down zero-mode is transported across the system in the negative $x$-direction. A schematic description of this process is presented in Fig.~\ref{xk_cylindrical}. 
This leads to $-1/2\pi$ for the Hall conductivity of spin-down zero-modes when $b_0 < 0$. 
Similar considerations imply that the zero modes contribute $+1/2\pi$ to $\sigma_{xy}^\uparrow$ when $b_0 > 0$.


 \begin{figure}[ht]
\includegraphics[width=3in]{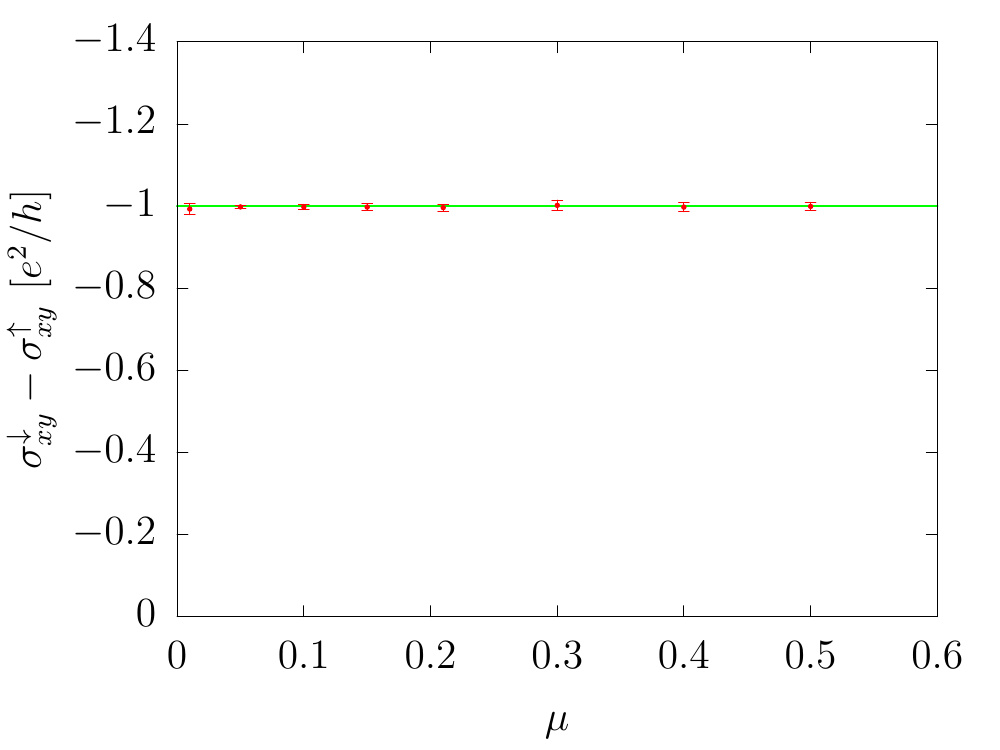}
\caption{Numerical calculation of $\sigma_{xy}^\downarrow-\sigma_{xy}^\uparrow$ for various Fermi energies for Hamiltonians in Eq.~\eqref{spinuphamiltonian} and~\eqref{spindownhamiltonian} for $b_0=0$. We have taken $m=1$. The energy and length scales are set by taking $k_F=1$ for the datapoint with highest Fermi energy. The cutoff in momentum space is $\Lambda = 3.3$ and system dimensions are $160\times 160$. In addition, the disorder strength and range defined in Eq.~\eqref{quencheddis} are $\sigma_V=0.034$ and $\mathcal{R}=6$. We have taken an average over $25$ disorder realizations and calculated standard devitiation to quantify the agreement between numerics and Eq.~\eqref{sumsecond} at $b_0=0$. It is shown as the errorbars in the figure which can be seen to be negligible. The horizontal green line represents $\sigma_{xy}^\downarrow - \sigma_{xy}^\uparrow = -1/2\pi$.}
\label{sumsecond_fig}
\end{figure} 

Having seen that Eq.~\eqref{sumsecond} applies quite generally for arbitrary $b(\bm r)$ such that $b_0 \neq 0$, it is natural to expect that it should be valid when $b_0 = 0$. 
To verify this, we have performed numerical simulations of the Hamiltonians in Eqs.~\eqref{susyone} and \eqref{susytwo}. The method closely follows Ref.~\onlinecite{2018arXiv180307767K}. 
Using a circular cutoff $\Lambda$ in the plane wave basis and a finite system size, the left-hand side of Eq.~\eqref{sumsecond} is calculated for various Fermi energies. 
We average over $N_{V} = 25$ disorder realizations and calculate the standard deviation (not the statistical error in the mean which would be smaller by a factor of $\sqrt{N_{V}}$). 
Our result is plotted in Fig.~\ref{sumsecond_fig}.
It agrees with the sum rule. 
Since the standard deviation is found to be negligible, Eq.~\eqref{sumsecond} holds to a very good accuracy for a single disorder realization. 
In addition, we find that the agreement improves upon increasing the system size and/or the disorder strength.

Based on the above discussion and the numerical calculation, we conclude that Eq.~\eqref{sumsecond} holds for arbitrary $b(\bm r)$ and $\mu > 0$. 
Along with Eq.~\eqref{sumfirst}, this establishes upon disorder averaging:
\begin{eqnarray}
\sigma_{xy}^{\rm (cf)} = -\frac{1}{4\pi}
\end{eqnarray}
at $\nu = 1/2$.

\subsection{HLR theory as a critical point:} 
\begin{figure}[ht]
\includegraphics[width=2.5in]{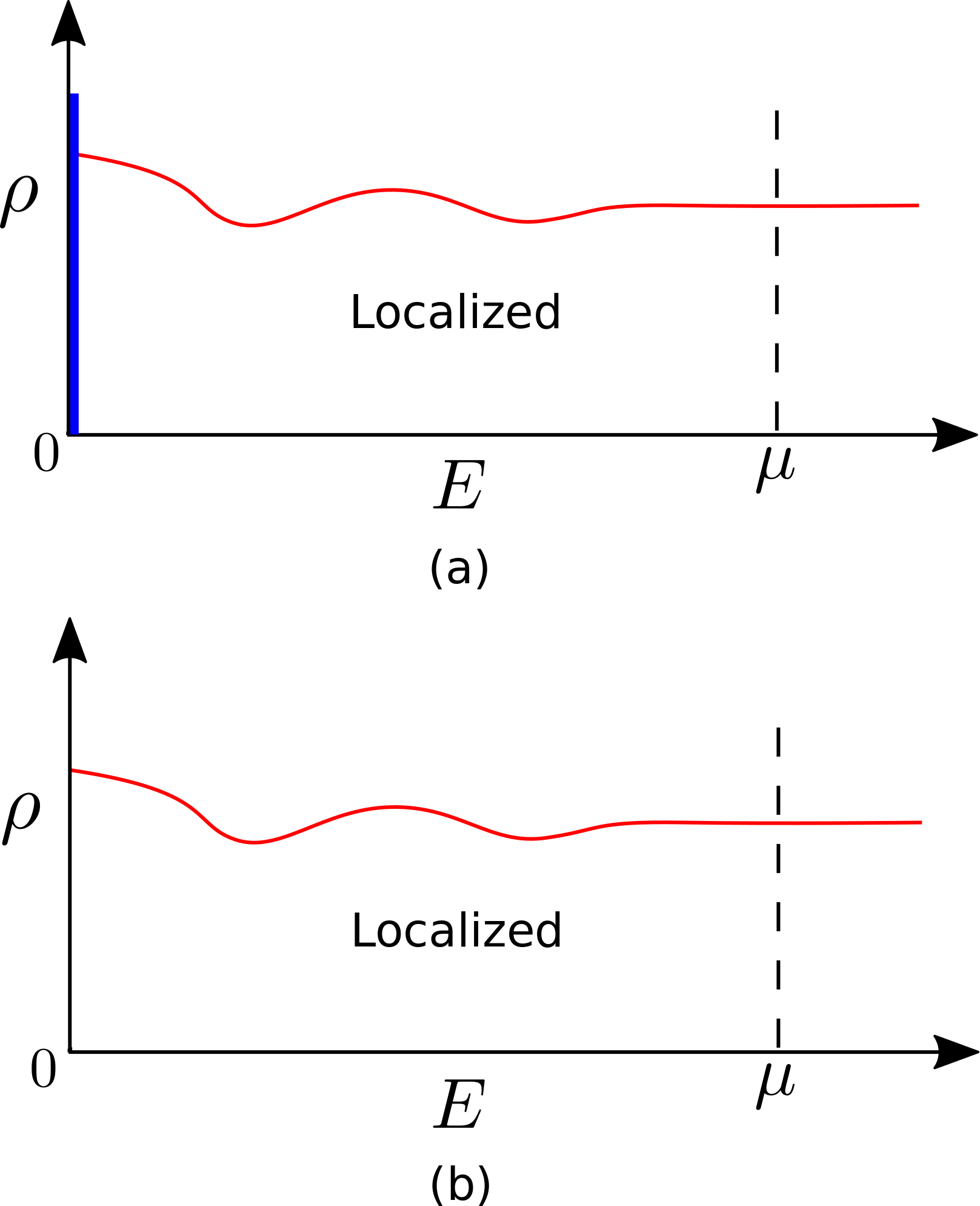}
\caption{A schematic for density of states of $\mathcal{H}_f$ in Eq.~\eqref{meanfieldH} for (a) $b_0 < 0 $ and (b) $b_0 > 0$ when spatial variations of magnetic field are large compared to $b_0$. Only (a) has zero-energy states that are shown as a Dirac delta function. The disorder washes out any quantum fluctuations in the DoS due to presence of a non-zero $b_0$ so that $\rho = m/2\pi$ at the Fermi-energy $\mu$. We assume that all extended states correspoding to higher Landau levels have levitated up\cite{PhysRevLett.52.2304} and thus all positive energy states below the Fermi-energy are localized.}
\label{DoS_strong_variations}
\end{figure}

In this section, we will argue that HLR theory lies at the critical point between two integer quantum Hall states. The effective magnetic field $b_0$ corresponds to an experimental tuning parameter that tunes the electronic system to $\nu=1/2$. Strictly speaking, a non-zero average magnetic field changes the compressibility of the system. However, we assume that $b_0$ is small so that the density of states (which is equal to the compressibility) gets smoothed out due to disorder and equals its value at $b_0=0$. 
A schematic DoS is shown in Fig.~\ref{DoS_strong_variations} for $b_0 <0$ and $b_0 > 0$.
As discussed in detail in the previous subsection, when $b_0 < 0$, the Hamiltonian $\mathcal{H}_f$ in Eq.~\eqref{meanfieldH} has zero-modes which contribute $-1/2\pi$ to the Hall conductivity. 
We assume that $b_0$ is small enough so that all other extended states corresponding to quantum Hall transitions in higher Landau levels have levitated up\cite{PhysRevLett.52.2304} in energy and all positive energy states are localized. Thus, the Hall conductivity at any Fermi energy is $\sigma_{xy}^{\rm (cf)} = -1/2\pi$. For the case $b_0>0$, there are no normalizable zero energy modes. Again, assuming that all positive energy states are localized, we find $\sigma_{xy}^{\rm (cf)} = 0$ for all Fermi energies. 
In summary,
\begin{eqnarray}
\sigma_{xy}^{\rm (cf)} =  \left\{ \begin{array}{cc} -\frac{1}{2 \pi}, &  b_0<0 \\
0, & b_0>0
\end{array} \right.
\end{eqnarray}
Thus, $b_0=0$ corresponds to the critical point of this integer quantum Hall transition of composite fermions. 

The tuning of effective magnetic field across $b_0 = 0$ can be understood as follows. When $b_0<0$, the number of zero-energy modes is given by $|b_0|L^2/2\pi$. Now, as we decrease $|b_0|$, the zero energy states levitate up one by one while keeping the Hall conductivity constant. For $b_0>0$, they have all levitated upward and the Hall conductivity is zero. This phase transition happens for all Fermi energies at exactly $b_0=0$ with $\sigma_{xy}^{\rm (cf)}=-1/4\pi$. Interestingly, this means that states at all Fermi energies become critical. This is consistent with the fact that changing the Fermi energy $\mu$ while keeping $b_0=0$ corresponds to changing the density of composite fermions and the external magnetic field $B$ in proportion to each other such that the electronic filling fraction $\nu=1/2$.

In terms of electrons, the dictionary between Hall conductivities of composite fermions and electrons implies that the state of electrons transitions from an integer quantum Hall state with $\sigma_{xy}=1$ to an insulator with $\sigma_{xy}=0$ as $b_0$ is tuned from negative to positive values. As a consequence, it is natural to identify the HLR theory as a critical theory for the $\nu=1$ to $\nu=0$ integer quantum Hall transition of electrons.

\section{Towards a low-energy effective theory}

In the previous section, we argued that the HLR mean-field theory exhibits
particle-hole symmetric Hall conductivity in the presence of weak, long-wavelength disorder.  
To conclude, we present a natural guess for a low-energy effective theory that produces such a response, upon disorder averaging.  
Provided that the range of the disorder $\mathcal R$ is large compared to the localization length, the simplest way to proceed is to spatially average Eq.~\eqref{effLagrangian1}: 
\begin{equation}
\label{effresponse}
\mathcal L_{\rm response,avg} = -\frac{1}{8 \pi} a d a.
\end{equation}
The $N$ dependence drops out after spatial averaging. (This term in the effective Lagrangian is also implied by the more detailed analysis of the previous section.)
The non-zero contribution arises from the ``unpaired" zero modes.  

However, as written, there is a problem with the Lagrangian in Eq.~\eqref{effresponse}: it violates gauge invariance\footnote{To be specific, the Lagrangian above is not invariant under large gauge transformations.}.\cite{RedlichKN, Redlichparitylong, AlvarezGaumeWitten1984}
The problem has a well-known resolution: in a simple-minded approach in which gauge fluctuations are neglected, one views the above Lagrangian as the result of integrating out a massive 2-component Dirac fermion (for consistency, we require that the mass of the fermion be greater than the Fermi energy $\mu$).  
In a well-defined theory with proper UV regularization (say on a lattice), there is a partner fermion which must remain gapless so that the effective theory describes a critical point: if it were massive, it would generate an order unity correction to the dc Hall conductance, which would violate the requirement of particle-hole symmetry.
Furthermore, the Lagrangian for the massless Dirac fermion must possess a statistical time-reversal symmetry (otherwise, it would contribute a non-zero correction to $\sigma^{\rm (cf)}_{xy} = - 1/4\pi$).  
Thus, a natural low-energy effective theory should take the form,\footnote{There could in principle be an additional even number of Dirac fermion modes that are massless.  However, such a situation generically would require fine-tuning, since the additional modes are not related to one another by any spatial symmetry}
\begin{equation}
\mathcal L_{\rm eff} = i \bar \psi \gamma_{\mu} D_{a}^{\mu} \psi 
-\frac{1}{8 \pi} a d a + \mathcal L_{\rm cs},
\end{equation}
where we have included the Chern-Simons term present in the original HLR Lagrangian in Eq.~\eqref{csterm}, which is a spectator for our mean-field analysis in which $a_\mu$ does not fluctuate. This term must be included in ${\cal L}_{\rm eff}$ to correctly translate composite fermion response to electrical response.
Simplifying the above expression we see that the CS terms for $a_{\mu}$ cancel and we arrive at the conjectured low-energy effective theory for a particle-hole symmetric Dirac composite fermion in Eq.~\eqref{Son}.

Further evidence for a massless Dirac Effective field theory comes from the fact that exactly at $b_0 = 0$, extended states occur at all energies.  
This is true for a single free 2-component massless Dirac fermion in the presence of pure flux disorder due to the presence of statistical time-reversal symmetry.

We have arrived at this theory by observing that slaving potential to flux disorder at $g=2$ realizes a critical point between an integer quantum Hall state ($\nu = -1$) and trivial insulator ($\nu = 0$).  
We stress that the above theory emerges only upon averaging over disorder in HLR theory and in a weak-coupling limit where gauge fluctuations have been neglected.  


To further motivate this conclusion, it is helpful to channel lattice models\cite{PhysRevLett.61.2015, PhysRevB.50.7526, 2006cond.mat..3271R} that realize integer quantum Hall transitions.   
In such models, fermions are subject to periodic chemical potentials as well as periodic magnetic fluxes (with net zero flux through the system).  At the critical point between integer quantum Hall and trivially insulating states, the low energy theory corresponds to a single 2-component Dirac fermion.  
Similar features appear to be at play here.  
In the present case, the Chern-Simons term in Eq.~\eqref{csterm} fixes the ratio of the potential and flux variations, effecting the tuning to a topological phase transition.   

We may speculate at this point on the effect of gauge fluctuations, which have been neglected throughout our analysis here.  
When gauge fluctuations are strong, the system is governed by a strong-coupling fixed point the nature of which remains poorly understood.  
However, following standard practice (see, e.g., Ref.~\onlinecite{NayakWilczek1994short}), gauge fluctuations can be treated perturbatively by including a variable-range Coulomb interaction between electrons.
As a result, we expect the theory above to hold, with some alterations which may include anomalous dimensions for various operators that smoothly vanish as the limit of a $1/|\bm r|$ Coulomb interaction is approached.\footnote{We expect fluctuations of the emergent Chern-Simons gauge field to generate, e.g., a random chemical potential and random mass, consistent with the broken time-reversal symmetry of each random vector potential realization. Nevertheless, a statistical time-reversal symmetry should be maintained in the presence of gauge fluctuations.}
For short-range gauge fluctuations, it remains unclear whether particle-hole symmetric response is preserved.

\section{Equivalent electrical response of HLR$^{g=2}$ and Dirac composite fermions}

In this section, we look at the response of HLR theory at $g=2$ to deviations from half-filling in the absence of disorder. We will substantiate the equivalence of the HLR theory at $g=2$ to the Dirac composite fermion theory by observing that they behave identically. 
Let us take the uniform effective magnetic field to be $b_0$. 
The Hamiltonian of composite fermions at $g=2$ is given by
\begin{eqnarray}
\mathcal{H} = \frac{(\bm p - \bm a)^2}{2m} + \frac{b_0}{2m},
\end{eqnarray}
where $\nabla \times \bm{a} = b_0$. If $b_0 > 0$, the $n^{th}$ Landau level has the energy $E_n = (n+1)\frac{|b_0|}{m}$ and if $b_0 < 0$, then the $n^{th}$ level is at energy $n\frac{|b_0|}{m}$. Working at constant chemical potential, we can see that if $p$ Landau levels are filled for $b_0 > 0$, $p+1$ Landau levels will be filled for $b_0 < 0$. Therefore, the response Lagrangian for these two cases is
\begin{align}
\mathcal{L}[a,A] = \zeta \frac{p+\frac{1-\zeta}{2}}{4\pi} ada + \frac{1}{8\pi} (a-A) d (a-A).
\end{align}
where $\zeta = {\rm sgn}(b_0)$. Upon integrating out the emergent gauge field $a$, we arrive at
\begin{eqnarray}
\mathcal{L}^{\rm eff}_{b_0>0} &=& \frac{1}{4\pi}\frac{p}{2p+1} A d A;\\
\mathcal{L}^{\rm eff}_{b_0<0} &=& \frac{1}{4\pi}\frac{p+1}{2p+1} Ad A .
\end{eqnarray}
Thus, we obtain particle-hole conjugate Jain sequence quantum Hall states quite naturally. 

We can compare this with the HLR theory at $g=0$, where the same number of Landau levels are filled for either sign of $b_0$. In this case, we would have been led to the filling fractions $\nu = \frac{p}{2p+1}$ and $\nu=\frac{p}{2p-1}$ that are not particle-hole conjugate of each other. 

Now, let us compare this with the response in Son's Dirac composite fermion theory.\cite{Son2015} To do this, we first rewrite Eq.~\eqref{Son} in the following manner:
\begin{align}
\mathcal L_{\rm Dirac} & = i \bar \psi \gamma_{\nu} D_a^{\nu} \psi + \frac{1}{8\pi} (a-A)d (a-A) - \frac{1}{8 \pi} a d a .
\,
\end{align}
This form allows for a direct comparison with the HLR theory. The last term above has the interpretation as the contribution from a massive Dirac partner whose mass is much larger than the energy scales of interest. For the remaining massless fermion, when we tune away from $\nu = 1/2$,
we fill $(p + 1/2)$ Landau levels for either sign of the effective magnetic field. Thus,
\begin{eqnarray}
\mathcal{L}[a,A] &=& \zeta \frac{p+\frac{1}{2}}{4\pi} ada + \frac{1}{8\pi}  (a-A)d(a-A) - \frac{1}{8 \pi}  a d a \nn\\
&=& \zeta \frac{p+\frac{1-\zeta}{2}}{4\pi} ada + \frac{1}{8\pi}  (a-A)d(a-A).
\end{eqnarray}
Consequently, the Dirac theory and the HLR theory with $g = 2$ have the same response properties. The massive partner effectively adds another half-filled Landau level that, in conjunction with the Landau levels of the massless Dirac fermion, precisely reproduces the non-relativistic spectrum of Landau levels at $g = 2$.

As a complementary case, we can also keep the electron density constant instead of the chemical potential. The HLR theory predicts that the same number of Landau levels would be filled for either sign of the effective magnetic field. Thus, the response and effective Lagrangians are
\begin{eqnarray}
\mathcal{L}[a,A] &=& \zeta \frac{p}{4\pi} ada + \frac{1}{8\pi}  (a-A)d(a-A); \\
\mathcal{L}^{\rm eff} &=& \frac{1}{4\pi}\frac{p}{2p+\zeta}AdA.
\end{eqnarray}
On the other hand, in the Dirac composite fermion theory, the composite fermion and electron densities are given by
\begin{eqnarray}
\rho_{\rm cf} &=& \frac{B}{4\pi};\\
\rho_e &=& \frac{B-b_0}{4\pi}.
\end{eqnarray}
Thus, we have
\begin{eqnarray}
\rho_{\rm cf} &=& \rho_e + \frac{b_0}{4\pi}.
\end{eqnarray}
Thus, composite fermion densities for the $\zeta= \pm 1 $ are
\begin{eqnarray}
\rho^{\pm}_{\rm cf} &=& \rho_e \pm \frac{|b_0|}{4\pi},\\
&{\rm so}&\nn\\
\rho^{+}_{\rm cf} - \rho^{-}_{\rm cf} &=& \frac{|b_0|}{2\pi}.
\end{eqnarray}
Since, the number of states in a Landau level are given by $|b_0|/2\pi$, we get that if $(p+1/2)$ Landau levels are filled for $b_0 >0$, then $(p-1/2)$ Landau levels are filled for $b_0 < 0$. Therefore, the response Lagrangian is
\begin{eqnarray}
\mathcal{L}[a,A] &=& \zeta \frac{p+\frac{\zeta}{2}}{4\pi} ada + \frac{1}{8\pi}  (a-A)d(a-A) - \frac{1}{8 \pi}  a d a \nn\\
 &=& \zeta \frac{p}{4\pi} ada + \frac{1}{8\pi}  (a-A)d(a-A).
\end{eqnarray}
We again find that the two theories produce identical response to the deviations from $\nu = 1/2$. However, we emphasize that the value of ``$g$'' in the HLR theory does not matter if one keeps the electron density constant.

\section{Discussion}

In the limit of vanishing disorder, Galilean invariance alone implies that a half-filled Landau level exhibits particle-hole symmetric electrical conductivity.\cite{1994cond.mat.10047M}
As such, when the zero-temperature dc longitudinal resistivity vanishes, both the Dirac\cite{Son2015} and HLR\cite{halperinleeread} composite fermion theories exhibit particle-hole symmetric response.
However, when spatial inhomogeneity is present, the issue of particle-hole symmetry is non-trivial.  
An appealing feature of the Dirac composite fermion theory is that it manifestly retains particle-hole symmetry in the presence of disorder at long distances.  
By contrast, as we have shown here, particle-hole symmetry in the HLR theory is a subtle emergent property that is only unearthed after a careful analysis of the problem, consistent with earlier work.\cite{PhysRevB.95.235424, 2017PhRvX...7c1029W, 2018arXiv180307767K} 

Our main conclusion is that 
the HLR theory exhibits particle-hole symmetric dc electrical response provided the spatial inhomogeneity preserves particle-hole symmetry upon disorder averaging (i.e., a statistical particle-hole symmetry).  
The key reason for this is that the HLR theory with quenched disorder, at least within a mean-field approximation where the fluctuations of the emergent Chern-Simons gauge field are ignored, is tuned to a topological quantum critical point at which the change in the composite fermion Hall conductivity $\Delta \sigma_{xy}^{\rm (cf)} = - e^2/h$.
The same is true of the Dirac theory with disorder, when analyzed to the same degree of approximation.  
If both theories describe a quantum critical point separating the same two phases, it is perhaps in keeping with notions of universality that both descriptions of the half-filled Landau level are the same at low energies.
 
The apparent (mean-field) equivalence of both the HLR and Dirac descriptions of the half-filled Landau level reminds us of network models for quantum Hall transitions.\cite{chalkercoddington}
 It is known that in such descriptions, the quantum Hall transition is a percolation transition, and emergent descriptions in terms of Dirac fermions can be obtained at the transition.  
 A related picture of a Dirac composite fermion theory arising at a percolation transition between the HLR composite fermion theory and its particle-hole conjugate\cite{BMF2015} was also found in Ref.~\onlinecite{mulliganraghufisher2016}.
 The extent to which such descriptions are related to the composite fermion theories studied here remains unclear at present.

 
We showed that the mean-field Hamiltonian governing HLR composite fermions is that of a free Fermi gas in slaved disorder (see Eq.~\eqref{slaving}) with gyromagnetic ratio $g=2$. 
The value $g=2$ is closely related to the fact that composite fermions at $\nu=1/2$ are thought of as electrons each bound to two units of flux.
It is natural to imagine that the HLR composite fermion theory for the state at $\nu=1/4$ can be similarly understood, but with a different gyromagnetic ratio corresponding to four flux quanta bound to each electron.
In our investigations, we have not found that such descriptions yield a composite fermion Hall conductivity equal to half the value at $\nu=1/2$. 
Because supersymmetry can be utilized only at $|g|=2$, there are no special properties expected away from this value, at least within our mean-field approach.
Nevertheless, it could be interesting to view the state at, say, $\nu=1/4$, hierarchically, i.e., as a metallic point that arises from composite fermions tuned to a composite fermion integer quantum Hall transition where $\Delta \sigma^{\rm (cf)}_{xy} = + e^2/h$. 

Our treatment of both theories within a mean-field approximation where gauge fluctuations are neglected is the reason that we only find a critical point, rather than a phase of matter.
Since free fermions with broken time-reversal symmetry undergo Anderson localization except at quantum critical points in two spatial dimensions, it follows that strong interaction effects are needed to describe a putative metallic phase in the vicinity of the half-filled Landau level.  While gauge fluctuations have certainly been studied in composite fermion theories, in the future, it will be interesting to study gauge fluctuation effects in the presence of quenched disorder.

Lastly, if both HLR and Dirac composite fermion descriptions are equivalent at long wavelengths, the extent to which their instabilities are identical remains unclear.\cite{Mross2015, 2016PhRvB..94k5304M, 2016PhRvB..94p5138W, 2018arXiv180310427B, 2018arXiv180401107M}  
Such instabilities occur both to incompressible fractional quantum Hall states as well as compressible anisotropic states and occur in half-filled higher Landau levels.  Assessing whether the different composite fermion descriptions of these phases are similar or distinct is a problem we wish to address in the future.  


\acknowledgments
We thank Jing-Yuan Chen and B. Halperin for illuminating discussions.
P.K. and  S.R are supported in part by the 
the DOE Office of Basic Energy Sciences, contract DE-AC02-
76SF00515.
M.M. is supported in part by the UCR Academic Senate and the Moore Foundation.
M.M. thanks the generous hospitality of the Stanford Institute for Theoretical Physics, where some of this work was performed.  

\appendix
\section{Proof of Eq.~\eqref{fermiindependence}\label{Efindependence}}
In this appendix, we will prove Eq.~\eqref{fermiindependence} using supersymmetric quantum mechanics and the Kubo formula.
Recall that the spin-up and spin-down Hamiltonians are
\begin{eqnarray}
\mathcal{H}_{\uparrow} &=& \frac{1}{2m} \left[(\bm{p}-\bm{a})^2 - b(\bm{r})\right],\\
\mathcal{H}_{\downarrow} &=& \frac{1}{2m} \left[(\bm{p}-\bm{a})^2 + b(\bm{r})\right].
\end{eqnarray}
As discussed in the main text, the positive energy states of the two Hamiltonians are related by $\ket{\psi_\uparrow} = \frac{Q}{\sqrt{E}}\ket{\psi_\downarrow}$ and the two Hamiltonians can be written using the operator $Q = \frac{\Pi_x -i \Pi_y}{\sqrt{2m}}$ as:
\begin{eqnarray}
\mathcal{H}_{\uparrow} &=& QQ^\dagger\nn ,\\
\mathcal{H}_{\downarrow} &=& Q^\dagger Q \nn .
\end{eqnarray}
The Hall conductivities of the two Hamiltonians can be written using the single particle Kubo-formula:
\begin{eqnarray}
\sigma_{xy}^\uparrow &=& -\frac{i 2\pi}{L^2} \sum_{n\neq l} f_n \frac{(v^\uparrow _x)_{nl}(v^\uparrow _y)_{ln} - (v^\uparrow _x)_{ln}(v^\uparrow _y)_{nl}}{(E_n-E_l)^2}\nn ,\\
\sigma_{xy}^\downarrow &=& -\frac{i 2\pi}{L^2} \sum_{n\neq l} f_n \frac{(v^\downarrow _x)_{nl}(v^\downarrow _y)_{ln} - (v^\downarrow _x)_{ln}(v^\downarrow _y)_{nl}}{(E_n-E_l)^2} \nn ,
\end{eqnarray}
where $f_n = \theta(\mu - E_n)$ is the Fermi-Dirac distribution at zero temperature for a Fermi energy $\mu$ and $v_i$ represents the velocity operator. 
Also, the indices $n,l$ run over the eigenstates of $\mathcal{H}_{\uparrow}$ in the first equation and the eigenstates of $\mathcal{H}_{\downarrow}$ in the second equation so that for positive energy states: $\ket{n_\uparrow} = \frac{Q}{\sqrt{E}}\ket{n_\downarrow}$. Now, 
\begin{eqnarray}
\hat{v}_x^\uparrow &=& \hat{v}_x^\downarrow = \frac{\Pi_x}{m} = \frac{Q^\dagger+Q}{\sqrt{2m}} \nn\\
\hat{v}_y^\uparrow &=& \hat{v}_y^\downarrow = \frac{\Pi_y}{m} = \frac{Q^\dagger-Q}{i\sqrt{2m}} \nn .
\end{eqnarray}
Using this, we can rewrite spin-up and spin-down Hall conductivities as
\begin{eqnarray}
\sigma_{xy}^\uparrow &=& \frac{ 2\pi}{L^2} \sum_{n\neq l} f_n \frac{|Q^\uparrow_{ln}|^2-|Q^\uparrow_{nl}|^2}{m(E_n-E_l)^2} \nn ,\\
\sigma_{xy}^\downarrow &=& \frac{2\pi}{L^2} \sum_{n\neq l} f_n \frac{|Q^\downarrow_{ln}|^2-|Q^\downarrow_{nl}|^2}{m(E_n-E_l)^2} \nn ,
\end{eqnarray}
where $Q^\uparrow_{ln} = \langle l_\uparrow|Q|n_\uparrow \rangle$ and similarly $Q^\downarrow_{ln} = \langle l_\downarrow|Q|n_\downarrow \rangle$.

Now take the derivative of both equations with respect to the Fermi energy $\mu$. 
Using $\frac{\partial f_n}{\partial \mu} = \delta (E_n-\mu)$ and considering $\mu > 0$, we have:
\begin{eqnarray}
\frac{\partial\sigma_{xy}^\uparrow}{\partial \mu} &=& \frac{2\pi}{L^2} \sum_{n\neq l} \delta(E_n-\mu) \frac{|Q^\uparrow_{ln}|^2-|Q^\uparrow_{nl}|^2}{m(E_n-E_l)^2} \label{d_sigma_xy_up}, \\
\frac{\partial\sigma_{xy}^\downarrow}{\partial \mu} &=& \frac{2\pi}{L^2} \sum_{n\neq l} \delta(E_n-\mu) \frac{|Q^\downarrow_{ln}|^2-|Q^\downarrow_{nl}|^2}{m(E_n-E_l)^2} .
\end{eqnarray}

We assume that $\frac{1}{L^2}\int d^2r\, b(\bm r) =b_0 > 0$ so that only the spin-up Hamiltonian has zero-energy modes. 
Using supersymmetric quantum mechanics, we have:
\begin{eqnarray}
Q^\downarrow_{ln} &=& \langle l_\downarrow|Q|n_\downarrow \rangle \nn\\
&=& \left(\langle l_\uparrow|\frac{Q}{\sqrt{E_l}}\right)Q \left(\frac{Q^\dagger}{\sqrt{E_n}}|n_\uparrow \rangle\right) \nn\\
&=& \frac{1}{\sqrt{E_n E_l}}\langle l_\uparrow|QQQ^\dagger |n_\uparrow \rangle \nn, \\
Q^\downarrow_{ln} &=& \sqrt{\frac{E_n}{E_l}}\langle l_\uparrow|Q|n_\uparrow \rangle = \sqrt{\frac{E_n}{E_l}}Q^\uparrow_{ln}.
\end{eqnarray}
Thus, the derivative of the spin-down Hall conductivity can be expressed in terms of spin-up eigenstates as
\begin{align}
\label{d_sigma_xy_down}
\frac{\partial\sigma_{xy}^\downarrow}{\partial \mu} =&\frac{ 2\pi}{L^2} \sum_{\substack{n\neq l \\ E_{l} \neq 0}} \delta (E_n-\mu) \frac{|Q^\uparrow_{ln}|^2 \frac{E_n}{E_l}- |Q^\uparrow_{nl}|^2\frac{E_l}{E_n}}{m(E_n-E_l)^2} .
\end{align}
Note that the sum over $n,l$ now goes over the positive energy eigenstates of $\mathcal{H}_\uparrow$. 

It is also useful to separate the contribution of zero-energy states in Eq,~\eqref{d_sigma_xy_up}:
\begin{eqnarray}
\frac{\partial\sigma_{xy}^\uparrow}{\partial \mu} &=&\frac{2\pi}{L^2} \sum_{\substack{n\neq l\\E_l \neq 0}} \delta (E_n-\mu) \frac{|Q^\uparrow_{ln}|^2- |Q^\uparrow_{nl}|^2}{m(E_n-E_l)^2}\nn\\
&\,&-\frac{2\pi}{L^2} \sum_{\substack{n\neq l\\E_l = 0}} \delta (E_n-\mu) \frac{|Q^\uparrow_{nl}|^2}{m E_n^2},
\label{d_sigma_xy_up2}
\end{eqnarray}
where we have used the fact that $Q^\dagger$ annihilates zero-energy states so that $Q^\uparrow_{ln}=\langle l_\uparrow|Q|n_\uparrow\rangle=0$ if $E_l=0$. 
Now subtract Eq.~\eqref{d_sigma_xy_down} from Eq.~\eqref{d_sigma_xy_up2}. 
After simplification, we find:
\begin{eqnarray}
\label{derivativeone}
&\,& \frac{\partial}{\partial \mu}(\sigma_{xy}^\downarrow -\sigma_{xy}^\uparrow) \nn\\
&\,& \,\,\,\,\,\,\,= \frac{2\pi}{L^2} \sum_{\substack{n\neq l \\ E_{l} \neq 0}} \delta (E_n - \mu) \frac{|Q^\uparrow_{ln}|^2 }{m E_l (E_n-E_l)}\nn\\
&\,& \,\,\,\,\,\,\, +\frac{ 2\pi}{L^2} \sum_{n\neq l} \delta (E_n-\mu) \frac{|Q^\uparrow_{nl}|^2 }{m E_n (E_n-E_l)}.
\end{eqnarray}
At this point, we can use the relation between the position operator and its time variation:
\begin{eqnarray}
i[\mathcal{H}_\uparrow,\hat{z}] = \frac{d\hat z}{dt} = \hat{v}_x + i \hat{v}_y = \sqrt{\frac{2}{m}}Q^\dagger
\end{eqnarray}
where $\hat{z} = \hat{x}+i \hat{y}$. This gives:
\begin{eqnarray}
\langle l_\uparrow |Q^\dagger| n_\uparrow \rangle &=& i\sqrt{\frac{m}{2}} \langle  l_\uparrow|[\mathcal{H}_\uparrow, \hat{z}]|n_\uparrow \rangle \nn \\
&=& i(E_l-E_n)\sqrt{\frac{m}{2}}\langle  l_\uparrow|\hat{z}|n_\uparrow \rangle
\end{eqnarray}
Substituting into Eq.~\eqref{derivativeone}, we obtain:
\begin{eqnarray}
&\,& \frac{\partial}{\partial \mu}(\sigma_{xy}^\downarrow -\sigma_{xy}^\uparrow) \nn\\
&\,& \,\,\,\,\,\,\,= \frac{i2\pi\sqrt{2m}}{L^2} \sum_{\substack{n\neq l \\ E_{l} \neq 0}} \delta (E_n-\mu) \frac{Q^\uparrow_{ln} z^\uparrow_{nl} }{E_l}\nn\\
&\,& \,\,\,\,\,\,\, -\frac{i 2\pi\sqrt{2m}}{L^2} \sum_{n\neq l} \delta (E_n-\mu) \frac{Q^\uparrow_{nl}z^\uparrow_{ln} }{E_n}.
\end{eqnarray}
We can now include the $n=l$ term and remove the complete set of states $\ket{l_\uparrow}\bra{l_\uparrow}$ to get:
\begin{eqnarray}
&\,& \frac{\partial}{\partial \mu}(\sigma_{xy}^\downarrow -\sigma_{xy}^\uparrow) \nn\\
&\,& \,\,\,\,\,\,\,= \frac{i2\pi\sqrt{2m}}{L^2} \sum_{n} \delta (E_n-\mu) \bra{n_\uparrow} \hat{z} P \frac{1}{\mathcal{H}_\uparrow}P Q\ket{n_\uparrow}\nn\\
&\,& \,\,\,\,\,\,\, -\frac{i 2\pi\sqrt{2m}}{L^2} \sum_{n} \delta (E_n-\mu) \bra{n_\uparrow}P \frac{1}{\mathcal{H}_\uparrow}P Q \hat{z}\ket{n_\uparrow}\nn,\\
&\,&
\end{eqnarray}
where $P$ projects onto the positive energy states of $\mathcal{H}_\uparrow$. Notice that $R = Q\frac{1}{\mathcal{H}_{\downarrow}}$ is a well defined operator since $\mathcal{H}_{\downarrow}$ does not have any eigenvalues equal to zero. Also, $R$ is the right multiplicative inverse of $Q^\dagger$, i.e., $Q^\dagger R = \mathbb{I}$, where $\mathbb{I}$ is the identity operator on the Hilbert space of ${\cal H}_\uparrow$.
It should be noted that $Q^\dagger$ does not have a left multiplicative inverse. 
Now insert $Q^\dagger R$ in appropriate places in the above scalar products:
\begin{eqnarray}
&\,& \frac{\partial}{\partial \mu}(\sigma_{xy}^\downarrow -\sigma_{xy}^\uparrow) \nn\\
&\,& \,\,\,\,\,\,\,= \frac{i2\pi\sqrt{2m}}{L^2} \sum_{n} \delta (E_n-\mu) \bra{n_\uparrow} \hat{z} P \frac{1}{\mathcal{H}_\uparrow}P QQ^\dagger R\ket{n_\uparrow}\nn\\
&\,& \,\,\,\,\,\,\, -\frac{i 2\pi\sqrt{2m}}{L^2} \sum_{n} \delta (E_n-\mu) \bra{n_\uparrow}P \frac{1}{\mathcal{H}_\uparrow}P QQ^\dagger R \hat{z}\ket{n_\uparrow}.\nn\\
&\,&
\end{eqnarray}
Since $QQ^\dagger = \mathcal{H}_\uparrow$ and $P \frac{1}{\mathcal{H}_\uparrow}P \mathcal{H}_\uparrow = P$:
\begin{eqnarray}
&\,& \frac{\partial}{\partial \mu}(\sigma_{xy}^\downarrow -\sigma_{xy}^\uparrow) \nn\\
&\,& \,\,\,\,\,\,\,= \frac{i2\pi\sqrt{2m}}{L^2} \sum_{n} \delta (E_n-\mu) \bra{n_\uparrow} \hat{z} PR-PR \hat{z}\ket{n_\uparrow} \nn.\\
&\,&
\end{eqnarray}
Since $Q^\dagger\ket{\psi} = 0$ for zero energy states, $R = Q\frac{1}{\mathcal{H}_\downarrow}$ projects the the bra multiplying on the left to positive energy states. Therefore, $PR = R$. Thus, we find:
\begin{eqnarray}
&\,& \frac{\partial}{\partial \mu}(\sigma_{xy}^\downarrow -\sigma_{xy}^\uparrow) \nn\\
&\,& \,\,\,\,\,\,\,= \frac{i2\pi\sqrt{2m}}{L^2} \sum_{n} \delta (E_n-\mu) \bra{n_\uparrow} [\hat{z},R]\ket{n_\uparrow} \label{Hall_cond_d_EF_intermediate}
\end{eqnarray}
We now use $[\hat{z},Q^\dagger] = 0$ to prove that $\langle n_\uparrow |[\hat{z},R]|n_\uparrow\rangle = 0$ for positive energy states.
To see this, we use a series of identities:
\begin{eqnarray}
\hat{z}Q^\dagger - Q^\dagger \hat{z} &=& 0\nn\\
\hat{z}-Q^\dagger \hat{z} R &=& 0\nn\\
Q^\dagger R \hat{z} - Q^\dagger \hat{z} R &=& 0 \nn\\
\bra{n_\downarrow} Q^\dagger [R,\hat{z}] \ket{n_\uparrow} &=& 0 \nn \\
\bra{n_\uparrow}[R,\hat{z}] \ket{n_\uparrow} &=& 0.
\end{eqnarray}
The last two steps are valid for positive energy states. Substituting into Eq.~\eqref{Hall_cond_d_EF_intermediate}, we find:
\begin{eqnarray}
\label{finalresult}
\frac{\partial}{\partial \mu}(\sigma_{xy}^\downarrow -\sigma_{xy}^\uparrow)=0
\end{eqnarray}
for $\mu > 0$ and when only $\mathcal{H}_\uparrow$ has zero-energy states. 
This is the case when $b_0 > 0$. 
When $b_0<0$, only $\mathcal{H}_\downarrow$ will have zero energy states. This derivation can then be repeated by exchanging the roles of up and down spins. Thus, Eq.~\eqref{finalresult} is valid for any $b_0 \neq 0$ and $\mu > 0$.
Notice that Eq.~\eqref{finalresult} holds for {\it each} disorder realization.

%
\bibliography{bigbib}
\bibliographystyle{utphys}

\end{document}